\documentclass[aps,prd,twocolumn,10 pt, superscriptaddress,nofootinbib]{revtex4-2}
\usepackage[utf8]{inputenc}
\usepackage[T1]{fontenc}
\usepackage{amsmath,amssymb}
\usepackage{graphicx}
\usepackage[normalem]{ulem}
\usepackage[colorlinks=true,linkcolor=blue,citecolor=blue,urlcolor=blue]{hyperref}

\newcommand{\K}{\mathcal{K}}
\newcommand{\Cth}{C_{\rm th}}
\newcommand{\OGW}{\Omega_{\rm GW}}
\newcommand{\betaG}{\beta_M^{\rm G}}
\newcommand{\dd}{\mathrm{d}}
\newcommand{\nbar}{\bar n}
   \newcommand{\nb}{\bar n}
\begin{document}

\title{Warm inflation beyond the Gaussian limit:%
\texorpdfstring{\\}{ }primordial black holes and induced gravitational waves}

\author{Mayukh R. Gangopadhyay}
\email{mayukhraj.g@vit.ac.in}
\affiliation{Department of Physics, School of Advanced Sciences, Vellore Institute of Technology (VIT), Chennai Campus, Chennai 600127, Tamil Nadu, India}

\author{Nilanjana Kumar}
\email{nilanjana.kumar@gmail.com}
\affiliation{Department of Physics, Chettinad Institute of Technology, Chettinad Academy of Research and Education, Chennai, Tamil Nadu 603103, India}

\begin{abstract}

In warm inflation, the primordial fluctuations are driven by random kicks from a hot radiation bath. Two features of these kicks matter: how long the bath remembers them, which shapes the power spectrum, and how their strengths are spread, which controls the rare, extreme fluctuations. In this paper, we investigate how primordial black holes and the accompanying induced gravitational waves can be used to discriminate between these two features. We use a simple model inspired by warm inflation, in which independent kicks of random strength add up to a narrow peak in the power spectrum. At leading order, the induced gravitational wave depends only on the power spectrum and is therefore insensitive to how the kicks are distributed. However, the black holes are very sensitive to these rare large fluctuations. Hence, two baths can yield the same gravitational wave signal while forming black holes at very different rates. A modest kurtosis range of 0.1 to 0.3 enhances black-hole formation by factors of ten to a million at fixed gravitational-wave amplitude, the larger factors for the rarer black holes. Counting peaks rather than all points, or widening the spectrum, lowers the enhancement by at most a factor of about four. In general a change in the power affects both phenomena, while the kurtosis affects only the black hole production.
\end{abstract}

\maketitle

\section{Introduction}
\label{sec:intro}

Primordial black holes (PBHs) form when rare, large curvature perturbations re-enter the horizon, and the same perturbations unavoidably produce a background of gravitational waves at second order \cite{ananda2007,kohri2018,espinosa2018,domenech2021}. Evidences of this induced background from pulsar timing arrays \cite{nanograv2023} and future interferometers \cite{lisa2017,et2020} are therefore routinely turned into statements about PBHs \cite{carr2020,sasaki2018}. That step relies on knowing the statistics of the curvature perturbation ($\zeta$). The induced waves respond mainly to the power spectrum, while PBHs come from the far tail of the distribution. Even a modest departure from Gaussianity changes the abundance of black holes by orders of magnitude \cite{byrnes2012,young2013,franciolini2018}. The exponential tails are produced by the quantum diffusion in stochastic inflation \cite{vennin2015,ezquiaga2020,pattison2021}.

Warm inflation \cite{berera1995,berera_fang1995,berera2009,bastero2009,kamali2023,moss2006,hall2004,basak2022} gives these statistics a physical origin. In the strong-dissipation regime \cite{gangopadhyay2021,bastero2016,berghaus2020,bastero2018,laine2021,berghaus2021,bastero2013} the inflaton perturbations are driven by the random force of a radiation bath. Several warm-inflation models produce enhanced small-scale power, PBHs and induced gravitational waves \cite{arya2020,bastero2021,ballesteros2022,correa2022,correa2024,gangopadhyay2026b}. The statistics of $\zeta$ are then inherited from the bath, and two of its properties come into play. The first is its \emph{memory}: a real bath has a finite correlation time, recorded in the covariance of the noise and the fluctuation-dissipation relation ties it to the dissipative response. In earlier work we showed how this memory reshapes the curvature spectrum \cite{gangopadhyay2026a}. The second is the \emph{statistics} of the force, that is how the strengths of the individual kicks are distributed. This information sits in the higher cumulants, on which the fluctuation-dissipation relation is silent. Matching that relation at the two point level does not make an arbitrary non-Gaussian jump process into a consistent thermal bath. The question we ask in the paper is whether PBHs and induced gravitational waves can discriminate these two properties: \emph{memory} and \emph{statistics}.

To address this issue we need a source in which these two properties will be calculable and they can be varied independently. We model the random force as a marked Poisson process: a stream of independent kicks whose strengths are drawn from a symmetric, bounded power law. The curvature responds linearly, through a kernel fixed by a narrow peak in the power spectrum. The model is motivated by warm inflation rather than derived from a particular potential, dissipation law or bath calculation, which would need the closed-time-path methods of Refs.~\cite{calzetta1988,boyanovsky1997,kamenev2011}. For the assumed Poisson source, the event density, mark distribution
and response kernel determine all cumulants. In this work, we fix the cumulant hierarchy such that it fixes all the higher even cumulants once the variance and the excess kurtosis are given, together with the shape of the mark distribution and the shape of the kernel.

The two observables respond to this source very differently. The power spectrum, and with it the leading induced-wave signal, depends only on the covariance of the force. Because the connected correlators of the model factorize into functions of the individual momenta, the angular selection rule of Garcia-Saenz \emph{et al.} \cite{garciasaenz2023} removes the trispectrum contribution at leading order. Zhou \emph{et al.}\cite{zhou2026} recently reached the same conclusion for the Poisson isocurvature left by evaporating PBHs. The induced waves are therefore blind to how the kicks are distributed. PBH formation is not. 

We compute the tail of the one-point distribution with saddle-point methods, which agree with an exact contour inversion out to twelve standard deviations, and feed it into a mass-weighted collapse statistic built on the compaction function \cite{musco2021,young2019}, critical collapse \cite{evans1994,niemeyer1998} and the mean profile of a rare peak. For the same induced-wave amplitude, a curvature kurtosis of $0.1$ to $0.3$ raises this statistic by a factor of $10^{0.9}$ to $10^{6.2}$; put the other way, a non-Gaussian bath can form the same black holes with induced waves up to about five times weaker. How large the effect is depends also on the response kernel and on the shape of the marks, and the model makes both dependences explicit.


We have performed a series of checks such as counting peaks instead of points, refining the collapse threshold and broadening the spectrum. We show that changing the response shifts the exponent by at most about $0.6$ and never removes the enhancement. Since our statistic is a one-point quantity rather than a full mass function, we use it mainly to compare baths with one another.

The sensitivity of PBHs to non-Gaussianity is well established. Our contribution has three parts. We give a source in which the covariance and the statistics can be varied independently, with every cumulant tied explicitly to the marks and the kernel. We compute the tail of the distribution accurately and check it against an exact calculation. And, within this model, we compare a collapse statistic directly with induced waves that respond only to the covariance. In this model, changing the covariance amplitude affects both the
collapse statistic and the leading induced-wave signal. At fixed
covariance, changing $\mathcal{K}$ affects the collapse statistic but
leaves the leading induced-wave signal unchanged. A joint measurement therefore carries information that neither observable carries alone, and can pick out the statistics of the source if the collapse threshold is known well; at our reference covariance a kurtosis of $0.1$ has the same effect as a $6$ per cent change in the threshold. It also follows that an upper limit on the induced background constrains PBHs only once the statistics of the source are fixed. A benchmark with asteroid-mass PBHs makes this concrete: the induced signal would lie far above the LISA sensitivity, and the challenge would be to interpret its amplitude rather than to detect it.

Section~\ref{sec:bath} sets up the source and the curvature field, and Sec.~\ref{sec:spectrum} discusses the spectrum and the induced waves. Section~\ref{sec:collapse} deals with the collapse statistic and the tail of the distribution. Section~\ref{sec:results} presents the results, Sec.~\ref{sec:robust} tests their robustness and Sec.~\ref{sec:consequences} discusses their consequences, including the physical benchmark. Section~\ref{sec:other} compares the model with other non-Gaussian constructions, and Sec.~\ref{sec:conclusions} concludes.

\section{The source and the curvature field}
\label{sec:bath}

We model the random force as a marked Poisson process. Events happen independently at positions $\mathbf{X}_j$, with a mean density $\nbar$ per unit volume once their time dependence has been integrated out (we return to this below). Each event carries a mark $x_j$, its signed strength, drawn from the normalised density
\begin{equation}
p(x)=\frac{\mu\,|x|^{-1-\mu}}{2\left(x_{\min}^{-\mu}-x_{\max}^{-\mu}\right)},\qquad x_{\min}<|x|<x_{\max}.
\label{eq:marks}
\end{equation}
This is a symmetric power law with $0<\mu<2$, cut off at both ends; Refs.~\cite{samorodnitsky1994,sato1999,applebaum2009,bouchaud1990,metzler2000,rosinski2007} give general background on jump processes of this kind. Our benchmark values are $\mu=1.5$ and $r\equiv x_{\min}/x_{\max}=10^{-3}$. Because of the cut-offs every moment of Eq.~(\ref{eq:marks}) is finite, and the condition $\mu<2$ simply means that the even moments are dominated by the strongest kicks, near $x_{\max}$. The field built from these bounded kicks is nevertheless unbounded: a Poisson sum can always stack up more kicks, so its tail reaches beyond any fixed multiple of $x_{\max}$. We will use the scaled moments $m_n=\langle x^n\rangle/x_{\max}^n$. For Eq.~(\ref{eq:marks}), $m_4/m_2=[(2-\mu)/(4-\mu)]\,(1-r^{4-\mu})/(1-r^{2-\mu})$, which is $0.207$ at the benchmark, while $m_4/m_2^2=2248$.

\begin{figure*}[t]
\centering
\includegraphics[width=\textwidth]{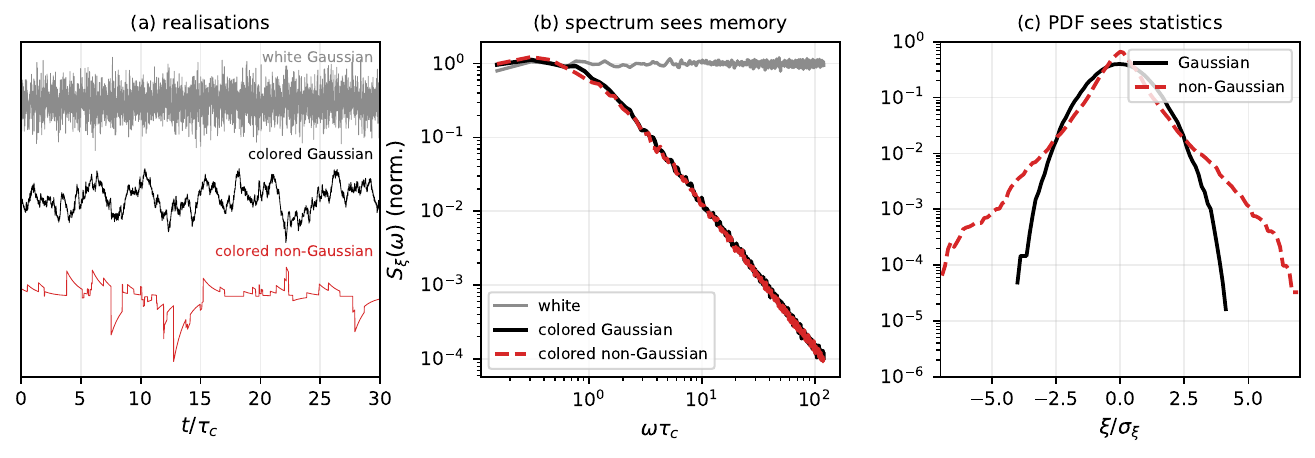}

\caption{\textit{Memory and statistics of a random force:} (a) Sample paths, each scaled to unit variance and shifted vertically for clarity: white Gaussian noise (top), Gaussian noise with correlation time $\tau_c$ (middle) and compound-Poisson noise with the same $\tau_c$ (bottom). The compound-Poisson force receives on average $\lambda\tau_c=2$ kicks per correlation time, with strengths drawn from Eq.~(\ref{eq:marks}) with $\mu=1.5$ and, for display, $r=0.05$; its variance rate is set equal to that of the Gaussian force, so the two forces with memory have the same covariance. (b) Their power spectra. Memory turns the flat white-noise spectrum into a Lorentzian, and the Gaussian and compound-Poisson spectra lie on top of each other, agreeing to within a few per cent. (c) Their one-point distributions: the spectra are the same, but the compound-Poisson force has much broader tails.}

\label{fig:force}
\end{figure*}

Figure~\ref{fig:force} shows the distinction we are after in the time domain. Start from white noise and give it a finite correlation time $\tau_c$: its flat spectrum becomes a Lorentzian, $S_\xi(\omega)\propto1/(1+\omega^2\tau_c^2)$. This is what we mean by memory, and it lives entirely in the covariance. Now keep the correlation time and the variance rate, but replace the Gaussian kicks by compound-Poisson kicks, so that the force becomes a jump-driven Ornstein--Uhlenbeck process \cite{marcus1978,chechkin2014,bns2001}. The spectrum does not change at all, yet the one-point distribution of the force develops broad tails. This is what we mean by statistics, and it lives in the higher cumulants. The rest of the paper follows each of these two properties through to the induced waves and to the black holes.

The curvature perturbation responds linearly to the events,
\begin{equation}
\zeta(\mathbf{x})=\sum_j x_j\,g\big(|\mathbf{x}-\mathbf{X}_j|\big),
\label{eq:field}
\end{equation}
where $g$ is an isotropic kernel. We use the convention $\langle\zeta_{\mathbf{k}}\zeta_{\mathbf{k}'}\rangle=(2\pi)^3\delta^3(\mathbf{k}+\mathbf{k}')P_\zeta(k)$, with dimensionless spectrum $\Delta^2_\zeta=k^3P_\zeta/(2\pi^2)$. The spectrum is then set by the Fourier transform $\tilde g(k)$ of the kernel, $P_\zeta(k)=\nb\langle x^2\rangle\,\tilde g(k)^2$. For a real isotropic kernel $\tilde g$ is real, so the spectrum fixes it only up to a sign at each $k$. We choose $\tilde g=+[P_\zeta/(\nb\langle x^2\rangle)]^{1/2}$ everywhere. Nothing in the spectrum forces this choice: a retarded mode response can oscillate and change sign, and other sign patterns give the same $P_\zeta$ with different higher statistics (Sec.~\ref{sec:robust} works through one example).

Campbell's theorem gives the cumulants of $\zeta(0)$ as
$\kappa_n=\bar n\langle x^n\rangle S_n$, where
$S_n=\int d^3r\,g(r)^n$. The variance is therefore
$\sigma^2=\bar n\langle x^2\rangle S_2$, and the excess kurtosis is:
\begin{equation}
\K=\frac{\kappa_4}{\sigma^4}=\frac{m_4}{m_2^2}\,\frac{S_4}{\nb\,S_2^2}.
\label{eq:kurtosis}
\end{equation}
Two features of this formula are worth noting. The mark factor is $m_4/m_2^2$, not the excess kurtosis of the marks themselves, and $\K$ is the kurtosis of $\zeta$, which is much smaller than that of a single kick because many events contribute at each point. All odd cumulants vanish, so the source itself introduces no skewness. (The nonlinear dynamics of warm inflation produces a bispectrum of its own \cite{moss2007,bastero2014, Mirbabayi2022}; we return to it in Secs.~\ref{sec:robust} and \ref{sec:other}.) The even cumulants follow a simple hierarchy,
\begin{equation}
\frac{\kappa_n}{\sigma^n}=\K^{\,n/2-1}\,\frac{m_n\,m_2^{\,n/2-2}}{m_4^{\,n/2-1}}\,R_n,\qquad
R_n=\frac{S_n\,S_2^{\,n/2-2}}{S_4^{\,n/2-1}},
\label{eq:hierarchy}
\end{equation}
for even $n\ge4$. Once the variance and $\K$ are fixed, the higher cumulants are set by the shape of the marks, through the $m_n$, and by the shape of the kernel, through the window factors $R_n$. If every event entered with equal weight, as for a set of identical cells, all $R_n$ would equal one, and $\K$ together with the marks would fix the whole distribution of $\zeta(0)/\sigma$. A real kernel does more. A three-dimensional Gaussian window gives $R_6=1.54$ and $R_8=2.83$, and the kernel of the narrow spectrum used below gives $R_6=2.71$ and $R_8=8.8$. The kernel therefore shapes the tail in its own right, even at fixed variance and kurtosis. 
The choice of Poisson may seem to be from the mathematical conviencence, but reason goes to a deeper physical attributes. In warm inflation the force on the inflaton comes from many separate microscopic processes in the bath, such as the sphaleron transitions of minimal warm inflation \cite{berghaus2020,laine2021}, each giving a short kick of its own size. When these kicks arrive independently and each one contributes only a small fraction of the total, the number that falls in any region follows Poisson statistics, and letting their sizes vary from kick to kick gives exactly the marked Poisson process of Eq.~(\ref{eq:marks}). Thus the choice of the Poisson kicks comes into this developement, the mathematical advantage of closed form cumulants is an added bonus.

So far each event has acted at a single instant. If the response separates into space and time, $g(r)f(t)$, integrating over the event times simply multiplies each mark by $f(t_j)$. A power law has no built-in scale, so the rescaled marks still follow $|x|^{-1-\mu}$ away from the cut-offs, which are smeared, and their moments factorise, $\langle(xf)^n\rangle=\langle x^n\rangle\langle f^n\rangle$. The cumulants become proportional to $\langle x^n\rangle S_nF_n$ with $F_n=\int\dd t\,f(t)^n$, so the time profile brings window factors of its own, $F_nF_2^{n/2-2}/F_4^{n/2-1}$, which multiply $R_n$. Our main results use a top-hat profile, for which these factors are one. An exponential profile multiplies $R_6$ and $R_8$ by $4/3$ and $2$, and Sec.~\ref{sec:robust} shows what this does. When the response does not separate, events at different times see different spatial kernels. The connected correlators of Sec.~\ref{sec:spectrum} still depend only on the magnitudes of the individual momenta, but the one-point statistics then depend on the full distribution of effective kick weights, a case we leave aside.

How large can $\K$ be in warm inflation? Equation~(\ref{eq:kurtosis}) has a simple reading, $\K=(m_4/m_2^2)/N_{\rm eff}$, where $N_{\rm eff}=\nb S_2^2/S_4$ is the effective number of events that feed $\zeta(0)$; for $N_c$ equal cells with $\nu$ events each, $N_{\rm eff}=\nu N_c$. In words, the kurtosis is the fourth-moment ratio of a single kick divided by the number of kicks. At the benchmark the mark factor is large, $m_4/m_2^2=2248$, because the bounded power law puts most of its fourth moment near the upper cut-off. Curvature kurtoses between $0.01$ and $0.3$ therefore need between $2.2\times10^5$ and $7.5\times10^3$ effective events. These are events, not bath cells. A rough count of the cells follows from a bath correlation length $\ell_{\rm corr}\sim1/T$ and the overdamped freeze-out wavenumber $k_s\simeq H\sqrt{3(1+Q)}$, with $Q=\Upsilon/3H$: the spatial part of the count is $(k_s\ell_{\rm corr})^{-3}$ \cite{graham2009,bastero2011}, and the temporal part depends on the dynamics of the bath \cite{gangopadhyay2026a}. For $Q=100$ and $T/H=30$ one finds $k_s\ell_{\rm corr}\simeq0.58$, or about five spatial cells. With a single event per cell the benchmark marks would give $\K\simeq440$, far beyond the range we consider. A kurtosis $\K=0.1$ needs some $4\times10^3$ events per cell, and the largest single kick is then comparable to the rms kick per cell: $y\equiv x_{\max}/(\nu\langle x^2\rangle)^{1/2}\simeq1.6$, with $\K=(m_4/m_2)\,y^2/N_c$. The fluctuation--dissipation relation fixes the covariance of the force and leaves the event rate and the mark law open, and this freedom is exactly what lets us vary the statistics at fixed covariance: we therefore treat $\K$ as a parameter between $10^{-2}$ and a few tenths, which captures both the event rate and the shape of the marks.

\section{The spectrum and the induced waves}
\label{sec:spectrum}

For a linear response, $\zeta_{\mathbf{k}}=\int\dd t\,G_k(t)\,\xi_{\mathbf{k}}(t)$, and the power spectrum is
\begin{equation}
P_\zeta(k)=\int\dd t\,\dd t'\,G_k(t)\,G_k^*(t')\,C_\xi(k;t,t').
\label{eq:Pzeta}
\end{equation}
It depends on the force only through its covariance $C_\xi$. Two baths with the same covariance and the same response therefore give the same $P_\zeta$, however different their higher cumulants may be; if the response changes as well, matching the covariance is no longer enough. In warm inflation the response also contains the growth factor that comes from the coupling between inflaton and radiation fluctuations \cite{graham2009,benetti2017}. This changes $P_\zeta$ but not the argument, which uses only linearity. Throughout we take a narrow log-normal peak,
\begin{equation}
\Delta^2_\zeta(k)=\frac{A}{\sqrt{2\pi}\,w}\exp\!\left[-\frac{\ln^2(k/k_p)}{2w^2}\right],\qquad w=0.1,
\label{eq:lognormal}
\end{equation}
normalised so that $\sigma^2=\langle\zeta(0)^2\rangle=\int\dd\ln k\,\Delta^2_\zeta=A$.

We compute the induced gravitational waves with the standard radiation-era kernel \cite{kohri2018}. Our implementation reproduces the known result $\OGW=0.8222\,A^2$ for a scale-invariant spectrum, and for Eq.~(\ref{eq:lognormal}) it gives a peak $\OGW\simeq1.87\,A^2$ at $k\simeq1.13\,k_p$ (Fig.~\ref{fig:profile}b). At fixed spectral shape the disconnected part of the signal therefore scales as $\OGW\propto A^2=\sigma^4$. Spectra with the same variance but different shapes would of course give different induced waves, and unless stated otherwise all statements below refer to the shape of Eq.~(\ref{eq:lognormal}).

What about the connected part, which is controlled by the trispectrum? Garcia-Saenz \emph{et al.} \cite{garciasaenz2023} showed that, once the two helicities are summed, the integrand carries an azimuthal factor $\cos2\psi$. Any trispectrum that does not depend on the crossed diagonals $|\mathbf{k}_1+\mathbf{k}_3|$ and $|\mathbf{k}_1+\mathbf{k}_4|$ therefore integrates to zero at leading order. Our field has exactly this property. For Eq.~(\ref{eq:field}), Campbell's theorem gives every connected correlator in closed form,
\begin{equation}
\langle\zeta_{\mathbf{k}_1}\cdots\zeta_{\mathbf{k}_N}\rangle_c=(2\pi)^3\delta^3\Big(\sum_i\mathbf{k}_i\Big)\,\nb\langle x^N\rangle\prod_{i=1}^N\tilde g(k_i),
\label{eq:campbell}
\end{equation}
a product of functions of the individual momentum magnitudes, with no dependence on how the momenta are combined. The selection rule then applies directly, and the connected contribution vanishes at leading order. Zhou \emph{et al.} \cite{zhou2026} recently used the same tensor projection and the same factorisation of Poisson correlators to show that connected terms drop out of the leading tensor power spectrum for the Poisson isocurvature left by evaporating PBHs; our result is the analogous statement for the marked bath. It concerns the leading tensor power spectrum of this family and rests on three natural ingredients: independent events, isotropic profiles and a linear response. Relaxing any of them lets the higher cumulants back in, as the detailed calculations for local-type models \cite{cai2019,adshead2021} and the treatments beyond a finite-order expansion \cite{iovino2024,zeng2025} show.

This leaves a clear division of roles between the two properties of the bath. Memory lives in the covariance, so a bath with a longer or shorter correlation time, of the kind studied in Ref.~\cite{gangopadhyay2026a}, changes the power spectrum and therefore shifts the induced gravitational waves and the black-hole abundance together. The statistics of the kicks act quite differently: they leave the power spectrum, and with it the leading induced waves, untouched, but they reshape the rare tail of the distribution from which black holes form. Within the model, then, the induced waves see only the covariance of the bath, while the black holes respond to both, and the rest of the paper is built on this difference.

\section{Collapse statistic and the tail}
\label{sec:collapse}


\begin{figure*}[t]
\centering
\includegraphics[width=\textwidth]{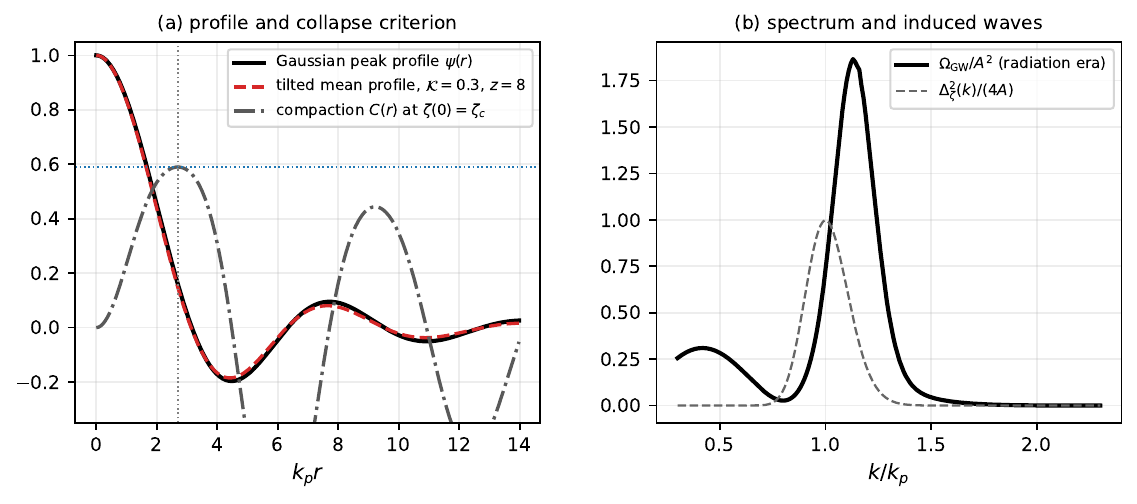}

\caption{(a) The shape of a rare peak and the collapse criterion: The solid black curve is the mean profile $\psi(r)$ of a rare peak in a Gaussian field. The red dashed curve is the mean profile of the tilted compound-Poisson ensemble, Eq.~(\ref{eq:profile}), for $\K=0.3$ and $\zeta(0)/\sigma=8$; it is barely distinguishable from $\psi$, with a slightly lower $c_1$ and a compaction maximum about two per cent further in. The grey dash-dotted curve is the compaction function of a peak exactly at the collapse threshold: its maximum, at $k_pr_m$ (vertical dotted line), just reaches $\Cth=0.59$ (horizontal dotted line). (b) The log-normal spectrum of Eq.~(\ref{eq:lognormal}), shown as $\Delta^2_\zeta/(4A)$ (dashed), and the gravitational waves it induces during radiation domination, $\OGW/A^2$ (solid), which peak at $k\simeq1.13\,k_p$.}

\label{fig:profile}
\end{figure*}

A region collapses to a black hole when the maximum of its compaction function exceeds a threshold \cite{musco2021}. For a spherically symmetric curvature profile $\zeta(r)$ on super-horizon scales the compaction function reads
\begin{equation}
C(r)=C_l(r)-\tfrac{3}{8}\,C_l(r)^2,\qquad C_l(r)=-\tfrac{4}{3}\,r\,\zeta'(r),
\label{eq:compaction}
\end{equation}

which already accounts for the nonlinear relation between the curvature and the density contrast \cite{young2019}. To see how this works, take a peak whose shape is fixed and only its height varies, $\zeta(r)=\zeta(0)\,\phi(r)$ with $\phi(0)=1$. In the type-I range \footnote{Type-I perturbations are those in which the areal radius $R=re^{\zeta(r)}$ still increases monotonically with $r$, which is the same as $C_l<4/3$, or $1+r\zeta'>0$, everywhere; larger, type-II perturbations fold the areal radius back on itself.} the compaction function peaks at the radius where $-r\phi'$ is largest, and this radius depends only on the shape, not on how tall the peak is. The maximum compaction is therefore a simple function of the height, $C_{\max}=c_1\zeta(0)-\frac{3}{8}c_1^2\zeta(0)^2$, where $c_1=\frac{4}{3}\max(-r\phi')$ is a single number that summarises the shape. Asking whether $C_{\max}$ exceeds $\Cth$ is then the same as asking whether the peak is tall enough, which gives an exact threshold on the amplitude,
\begin{equation}
\zeta(0)>\zeta_c=\frac{4}{3c_1}\Big[1-\big(1-\tfrac{3}{2}\Cth\big)^{1/2}\Big],
\label{eq:zetac}
\end{equation}
valid as long as $\zeta(0)<4/(3c_1)$, the type-I limit. For a rare peak of a Gaussian field the mean profile is the normalised correlation function $\psi(r)$ \cite{bbks1986}. For the spectrum of Eq.~(\ref{eq:lognormal}) this gives $k_pr_m=2.70$ and $c_1=1.379$. With $\Cth=0.59$, the value found for a narrow spectrum when nonlinear effects are included \cite{musco2021}, the threshold is $\zeta_c=0.639$, comfortably below the type-I limit of $0.967$. The shape prescription of Ref.~\cite{musco2021} applied to $\psi(r)$ itself gives $\Cth\simeq0.593$ (Sec.~\ref{sec:robust}), consistent with this choice.

A change in the statistics can also change the shape of a rare peak. For the field of Eq.~(\ref{eq:field}) we estimate the profile around a point with given $\zeta(0)$ by the mean of the Poisson ensemble exponentially tilted so that $\langle\zeta(0)\rangle$ equals that value, taken at the saddle point $\hat s$ of the tail calculation described below \cite{dembo1998}:
\begin{equation}
\langle\zeta(\mathbf{r})\rangle_{\hat s}=\nbar\int\dd^3y\;g(|\mathbf{r}-\mathbf{y}|)\,\Big\langle x\,\sinh\!\big(\hat s\,x\,g(y)/\sigma\big)\Big\rangle .
\label{eq:profile}
\end{equation}
Equation~(\ref{eq:profile}) is the expectation in the tilted ensemble, which approximates the conditional mean profile when $\zeta(0)$ is large. It should not be confused with an exact conditional peak profile, and a point with a high central value need not have the profile that crosses the threshold most easily. For a Gaussian field Eq.~(\ref{eq:profile}) reduces to $\zeta(0)\psi(r)$, as it should. For the compound-Poisson field the profile changes slightly: $c_1$ decreases by $0.2$--$1.8$ per cent for $\K$ between $0.01$ and $0.3$ and $\zeta(0)/\sigma$ between $6$ and $10$, and the maximum of the compaction function moves inward by up to about two per cent (Fig.~\ref{fig:profile}a). Since a lower $c_1$ raises the amplitude required for collapse, we use at each amplitude the value of $c_1$ of the corresponding tilted profile, keeping $\Cth=0.59$. In this section and the next we neglect fluctuations about the mean profile and count points instead of maxima. Section~\ref{sec:robust} estimates both effects and also recomputes the threshold for each tilted profile.

The mass of the black hole obeys critical scaling, $M=c_{\rm cr}M_H\,(C_{\max}-\Cth)^\gamma$ with $\gamma\simeq0.36$ \cite{evans1994,niemeyer1998}, and we define the mass-weighted collapse statistic
\begin{equation}
\beta_M=\int_{\zeta_c}^{\zeta_{\rm I}}\dd\zeta_0\;p(\zeta_0)\;c_{\rm cr}\big[C_{\max}(\zeta_0)-\Cth\big]^\gamma ,
\label{eq:beta}
\end{equation}
with $c_{\rm cr}=1$ throughout. Because $c_1$ depends on $\zeta_0$ the limits are implicit: the lower one solves $\zeta_0\,c_1(\zeta_0)=\frac{4}{3}[1-(1-\frac{3}{2}\Cth)^{1/2}]$, and the upper one, the type-I limit, solves $\zeta_0\,c_1(\zeta_0)=4/3$. The statistic differs from the bare probability $P(\zeta(0)>\zeta_c)$ through the critical-collapse weight. It is a one-point quantity: it counts points rather than maxima, and it integrates over the masses formed at a single formation scale, so it should not be read as a differential abundance at fixed PBH mass, which would require a mass-function calculation. The constant $c_{\rm cr}$ and the conversion from $\beta_M$ to a present-day abundance at the same formation epoch cancel in ratios at fixed covariance, although the value of $c_{\rm cr}$ affects the covariance selected by an absolute reference value $\betaG$, and hence the numbers quoted below. 

We include no peak-volume factor, and we let the linear map of Eq.~(\ref{eq:field}) carry the field all the way up to the threshold, as is natural in a model built on a linear response. Because $\zeta_c\simeq0.64$ is not small, Sec.~\ref{sec:robust} also follows what happens when we add a quadratic correction.

To evaluate Eq.~(11), we now calculate the probability density of the
central curvature amplitude. For $Z=\zeta(0)/\sigma$ the cumulant generating function is known in closed form,
\begin{equation}
\mathcal{C}(s)=\nbar\int\dd^3r\,\Big(\big\langle\cosh\big(s\,x\,g(r)/\sigma\big)\big\rangle-1\Big),
\label{eq:cgf}
\end{equation}
so the tail can be computed directly instead of being pieced together from a few moments. We evaluate the tail probability with the Lugannani--Rice formula \cite{daniels1954,lugannani1980},
\begin{equation}
P(Z>z)\simeq1-\Phi(w)+\varphi(w)\Big(\frac{1}{u}-\frac{1}{w}\Big),
\label{eq:LR}
\end{equation}

where $\hat s$ solves $\mathcal{C}'(\hat s)=z$, $w=[2(\hat sz-\mathcal{C}(\hat s))]^{1/2}$ and $u=\hat s\,\mathcal{C}''(\hat s)^{1/2}$, and the density with the corresponding saddle-point expression $p(z)\simeq e^{\mathcal{C}(\hat s)-\hat sz}/[2\pi\mathcal{C}''(\hat s)]^{1/2}$. Both approximations are very accurate here. Compared with a numerically converged inversion along a contour through the saddle point, for equal weights and for the kernel of Eq.~(\ref{eq:lognormal}), out to $\zeta(0)/\sigma=12.3$, the tails agree to within $0.8$ per cent and the densities to within $1.9$ per cent. The one-point distribution is therefore well under control, and the remaining questions concern the formation prescription itself, which we take up in Sec.~\ref{sec:robust}.


\section{Results}
\label{sec:results}

\begin{figure*}[t]
\centering
\includegraphics[width=\textwidth]{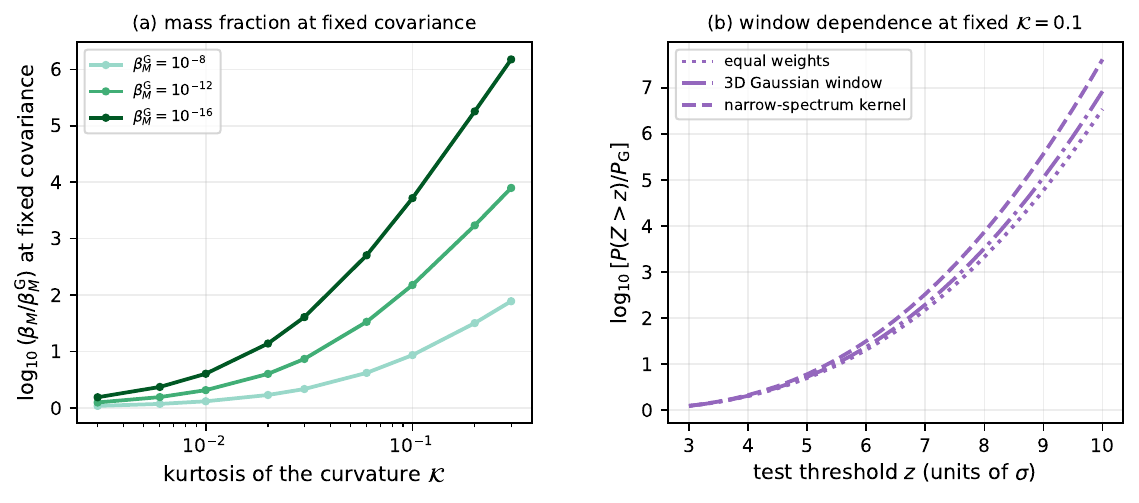}
\caption{(a) How much a non-Gaussian bath enhances the collapse statistic over a Gaussian bath with the same covariance, and hence the same induced-wave amplitude, as a function of the curvature kurtosis. The three curves correspond to the three covariances of Table~\ref{tab:enhancement}, labelled by their Gaussian values. (b) The role of the response kernel. At fixed $\K=0.1$, the enhancement of the one-point tail is plotted against a test threshold $z$ in units of $\sigma$, distinct from the collapse threshold $\zeta_c$, for equal weights, for a three-dimensional Gaussian window and for the kernel of the narrow log-normal spectrum. The kernel with the largest window factors gives the heaviest tail.}

\label{fig:enhancement}
\end{figure*}

Table~\ref{tab:enhancement} and Fig.~\ref{fig:enhancement}a show the collapse statistic of the compound-Poisson bath relative to that of a Gaussian bath with the same covariance, for three values of the covariance which we label by the Gaussian value $\betaG$. The three cases correspond to $\sigma=0.1207$, $0.0946$ and $0.0802$, or $\zeta_c/\sigma=5.29$, $6.76$ and $7.97$. To take one entry: at $\betaG=10^{-12}$, a kurtosis $\K=0.1$ raises the statistic by a factor $10^{2.18}$. Had we kept the Gaussian profile instead of the tilted one, the factor would have been $10^{2.31}$, so the change in the profile reduces the enhancement somewhat but is far from removing it.

\begin{table}[t]
\caption{Enhancement of the collapse statistic, $\log_{10}(\beta_M/\betaG)$, for a compound-Poisson bath relative to a Gaussian bath with the same covariance, and hence the same induced-wave amplitude. Each row is one curvature kurtosis, and each column one covariance, labelled by the value $\betaG$ that the Gaussian bath gives there. All entries use the spectrum of Eq.~(\ref{eq:lognormal}), the benchmark marks, the compaction criterion with $\Cth=0.59$, critical-collapse weighting with $c_{\rm cr}=1$ and the tilted mean profile of Eq.~(\ref{eq:profile}). Table~\ref{tab:robust} shows how these numbers change when the collapse statistic is refined.}

\label{tab:enhancement}
\begin{ruledtabular}
\begin{tabular}{lccc}
 & $\betaG=10^{-8}$ & $10^{-12}$ & $10^{-16}$ \\
\hline
$\K=0.01$ & 0.12 & 0.32 & 0.61 \\
$\K=0.03$ & 0.33 & 0.87 & 1.61 \\
$\K=0.1$  & 0.94 & 2.18 & 3.72 \\
$\K=0.3$  & 1.89 & 3.90 & 6.18 \\
\end{tabular}
\end{ruledtabular}
\end{table}
\begin{table}[t]
\caption{How the response kernel shapes the tail: The entries are $\log_{10}[P(Z>z)/P_{\rm G}(Z>z)]$, the enhancement of the one-point tail over a Gaussian with the same variance, at fixed kurtosis, for the benchmark marks and three kernels: equal weights, a three-dimensional Gaussian window and the kernel of the narrow spectrum. Here $z$ is a test threshold in units of $\sigma$, not the collapse threshold. The kernel with the largest window factors $R_n$ gives the heaviest tail.}

\label{tab:kernels}
\begin{ruledtabular}
\begin{tabular}{ccccc}
$\K$ & $z$ & equal weights & 3D Gaussian & narrow-spectrum kernel \\
\hline
0.1 & 7 & 2.179 & 2.293 & 2.521 \\
0.1 & 8 & 3.326 & 3.513 & 3.869 \\
0.3 & 8 & 5.459 & 5.770 & 6.278 \\
\end{tabular}
\end{ruledtabular}
\end{table}

The size of the enhancement also depends on the kernel, as Table~\ref{tab:kernels} and Fig.~\ref{fig:enhancement}b show: at fixed variance and fixed $\K$ they compare the one-point tail obtained with equal weights, with a three-dimensional Gaussian window and with the kernel of the narrow spectrum, as a function of a test threshold $z=\zeta(0)/\sigma$. The larger $R_n$ of the kernel make its tail heavier, by a factor of $2.2$ relative to equal weights at $\K=0.1$ and $z=7$ and by a factor of $6.6$ at $\K=0.3$ and $z=8$. It depends on the shape of the marks as well: at $\K=0.1$ and $\betaG=10^{-12}$, $\log_{10}(\beta_M/\betaG)$ is $2.61$ for $\mu=1.9$ and $2.09$ for $r=0.05$, compared with $2.18$ at the benchmark, i.e.\ changes in the statistic by factors of $2.68$ and $0.81$. These two cases illustrate the size of the effect, and we have not scanned the full space of admissible mark distributions.


\section{Robustness of the collapse statistic}
\label{sec:robust}

The results of Sec.~\ref{sec:results} rest on a few approximations in the collapse statistic and on a few choices in the model. Table~\ref{tab:robust} varies them one at a time and shows how the enhancement at the reference $\betaG=10^{-12}$ responds. In every case it stays large.

\begin{table}[t]
\caption{Robustness of the enhancement $\log_{10}(\beta_M/\betaG)$ at $\betaG=10^{-12}$. Each row changes one ingredient relative to the baseline of Table~\ref{tab:enhancement}: rows 2--5 refine the collapse statistic, and rows 6--11 change the model. The rows for $w=0.2$ and $0.3$ use the shape-dependent threshold for every profile and are measured against their own Gaussian reference. The enhancement remains large in every case.}
\label{tab:robust}
\begin{ruledtabular}
\begin{tabular}{lcc}
 & $\K=0.1$ & $\K=0.3$ \\
\hline
baseline (Table~\ref{tab:enhancement}) & 2.18 & 3.90 \\
shape-dependent threshold & 2.22 & 3.97 \\
profile scatter & 2.17 & 3.88 \\
maxima instead of points & 1.90 & 3.41 \\
scatter and maxima & 1.89 & 3.39 \\
exponential time response & 2.35 & 4.17 \\
kernel sign flipped below $k_p$ & 2.10 & 3.94 \\
quadratic map, $f_{\rm NL}=-0.5$ & 2.13 & 3.84 \\
quadratic map, $f_{\rm NL}=+0.5$ & 2.22 & 3.94 \\
width $w=0.2$ & 1.89 & 3.46 \\
width $w=0.3$ & 1.76 & 3.32 \\
\end{tabular}
\end{ruledtabular}
\end{table}

The first approximation concerns the threshold. Holding $\Cth=0.59$ while $c_1$ changes, as we did above, treats a change in the shape of the profile as if it were only a change in its height. Musco \emph{et al.} \cite{musco2021} give a threshold for a profile of arbitrary shape: a shape parameter $\alpha$, read off from the profile at $r_m$, is mapped to the threshold through the analytic form of Escriv\`a \emph{et al.} \cite{escriva2020}. Our implementation reproduces their monochromatic case, with $k_pr_m=2.74$, $\alpha=6.33$ and $\Cth\simeq0.59$, and for $\psi(r)$ it gives $\alpha\simeq6.1$ and $\Cth\simeq0.593$. We keep $\Cth=0.59$ for $\psi$ and rescale the threshold of each tilted profile by the ratio of the two prescribed values. The tilted profiles have a slightly smaller $\alpha$, which lowers their threshold by $0.1$--$0.4$ per cent and raises the enhancement by at most $0.11$ in $\log_{10}$ anywhere in Table~\ref{tab:enhancement}. Keeping the threshold fixed therefore slightly understates the enhancement, so on this count the numbers of Table~\ref{tab:enhancement} are, if anything, on the low side.

The second approximation is to ignore the scatter of the profile about its mean. The tilted ensemble is still a Poisson field, with the mark density reweighted by $e^{\hat s x g(y)/\sigma}$, so its covariance is known exactly. Treating the fluctuations about the mean as Gaussian and conditioning on $\zeta(0)$, we sample spherically averaged profiles, find $C_{\max}$ for each and average the critical-collapse weight. Some regions slightly below $\zeta_c$ now collapse and some slightly above it do not, but for a narrow spectrum the spherical average is almost fixed by $\zeta(0)$, and the enhancement moves by $0.02$ or less at all three reference covariances.

The third approximation, counting points instead of maxima, has a larger effect. Following Ref.~\cite{bbks1986}, we weight each height by the density of maxima there: the probability density of a vanishing gradient times $\langle|\det H|\,\Theta(-H)\rangle$, both evaluated with the tilted covariance of the gradient and the Hessian $H$ at the origin. Rare kicks raise the variance of the local gradient in the tilted ensemble, so a point of given height is somewhat less likely to be a maximum than in the Gaussian field; at the same height the density of maxima is lower by $0.25$--$0.7$ in $\log_{10}$ for $\K$ between $0.1$ and $0.3$. Counting maxima therefore lowers $\log_{10}(\beta_M/\betaG)$, at $\K=0.1$ by $0.21$, $0.28$ and $0.34$ for $\betaG=10^{-8}$, $10^{-12}$ and $10^{-16}$, and at $\K=0.3$ by $0.39$, $0.49$ and $0.57$. This is the largest correction we find. It trims the one-point numbers of Table~\ref{tab:enhancement} by up to about half a unit in the exponent and leaves the enhancement firmly in place.

The remaining rows of Table~\ref{tab:robust} change the model itself. A spectrum three times broader, $w=0.3$, has smaller window factors, $R_6=1.65$ and $R_8=3.28$, and its tilted profiles change more, with $c_1$ falling by $6$--$8$ per cent at the threshold. With the shape-dependent threshold applied throughout (it gives $\Cth=0.569$ for the Gaussian profile of this spectrum), the enhancement goes from $10^{2.22}$ to $10^{1.76}$ at $\K=0.1$ and from $10^{3.97}$ to $10^{3.32}$ at $\K=0.3$, with $w=0.2$ in between. The spectral shape thus matters at the level of half a unit in the exponent, another sign that $\K$ alone does not fix the answer. An exponential time profile of the response pushes the other way: it raises the window factors, and with them the enhancement, by $0.17$ and $0.27$ in the exponent. Flipping the sign of $\tilde g$ below $k_p$ is a more subtle test. It leaves the spectrum unchanged, but the jump at $k_p$ makes the kernel fall off only as $\cos(k_pr)/r^2$, so its radial integrals converge slowly; a cut at $k_pr=90$, ample for the positive kernel, would miss $5.7$ per cent of $S_2$. Carried out to $k_pr=2880$, where $S_2$ agrees with its value from Parseval's theorem to $0.03$ per cent, the integrals give $R_6=2.56$ and $R_8=9.10$, close to the values $2.71$ and $8.8$ of the positive kernel, and the enhancement changes by only $-0.08$ at $\K=0.1$ and $+0.04$ at $\K=0.3$. Finally, the field need not stay exactly linear on its way to the threshold. The nonlinear dynamics of warm inflation would add a quadratic piece \cite{moss2007,bastero2014}, which we model as $\zeta=\zeta_L+\frac{3}{5}f_{\rm NL}\zeta_L^2$, applied to the Gaussian reference and to the Poisson field alike. For $f_{\rm NL}=\pm0.5$ the enhancement at fixed $\betaG$ stays within $0.05$ of its linear value, because the quadratic term shifts the threshold of the two fields in the same way. What it does change is the covariance needed to reach a given $\betaG$, by an amount equivalent to a factor $1.23$ ($f_{\rm NL}=-0.5$) or $0.76$ ($f_{\rm NL}=+0.5$) in the leading induced-wave amplitude. In the plane of Sec.~\ref{sec:separating} a nonlinearity of this kind therefore looks like a change in the covariance amplitude rather than a change in $\K$, and we return to it there.

\section{Consequences}
\label{sec:consequences}

\subsection{The collapse statistic at a given induced-wave amplitude}

Because the leading induced waves depend only on the covariance, Table~\ref{tab:enhancement} can be read in a second way: as the range of collapse statistics that are all compatible with the same induced-wave amplitude, for this spectrum. If the curvature kurtosis is uncertain at the level of $0.1$, the statistic inferred from a measured amplitude is uncertain by a factor between $10^{0.9}$ and $10^{3.7}$ over the covariances of the table; at the level of $0.3$ the factor grows to between $10^{1.9}$ and $10^{6.2}$. Counting maxima trims these exponents by the amounts given in Sec.~\ref{sec:robust}, but the spread remains large. An upper limit on the induced background therefore becomes an upper limit on the black-hole abundance only once the statistics of the source have been fixed.

\begin{figure}[t]
\centering
\includegraphics[width=\columnwidth]{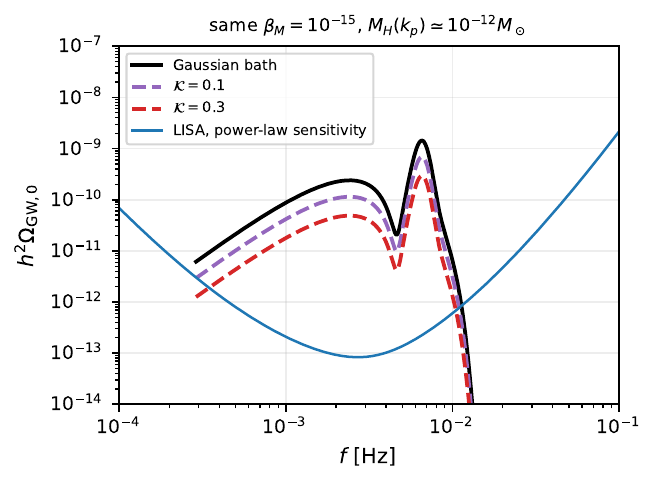}
\caption{Physical benchmark. The induced gravitational-wave background today for $k_p\simeq3.7\times10^{12}\,{\rm Mpc}^{-1}$, the scale that re-enters the horizon when $M_H\simeq10^{-12}M_\odot$. The curves show a Gaussian bath (solid) and compound-Poisson baths with $\K=0.1$ (purple, dashed) and $\K=0.3$ (red, dashed), all tuned to the same collapse statistic $\beta_M=10^{-15}$, which corresponds to $f_{\rm PBH}\simeq0.4$ when read as a formation mass fraction. The non-Gaussian baths reach this value with induced waves about two and five times weaker. The blue curve is the LISA power-law-integrated sensitivity for a signal-to-noise ratio of $10$ in four years, built from the instrument noise of Ref.~\cite{robson2019} with the method of Ref.~\cite{thrane2013}.}

\label{fig:benchmark}
\end{figure}

It helps to put these numbers on a physical scale. Suppose $k_p$ re-enters the horizon when the horizon mass is $M_H\simeq10^{-12}M_\odot$, in the asteroid-mass window where PBHs could make up all of the dark matter. With $g_*=106.75$ this means $k_p\simeq3.7\times10^{12}\,{\rm Mpc}^{-1}$, the induced waves peak at $f\simeq6.5$~mHz, in the LISA band, and today $h^2\Omega_{{\rm GW},0}\simeq1.6\times10^{-5}\,\OGW$ \cite{domenech2021}. The benchmark also needs a sensible value of the collapse statistic. Reading $\beta_M$ as the formation mass fraction, the standard conversion \cite{carr2021}, which depends only on the horizon mass at formation, gives $f_{\rm PBH}=1$ at $\beta\simeq2.6\times10^{-15}$, so the reference value $10^{-12}$ used in the tables would overproduce the dark matter by a factor of about $400$ at this mass. We therefore set $\beta_M=10^{-15}$, which corresponds to $f_{\rm PBH}\simeq0.4$. Since $\beta_M$ is a one-point statistic with $c_{\rm cr}=1$, this calibration is an order-of-magnitude guide that places the benchmark in an allowed range, and it plays no role elsewhere in the paper. For the three baths that give $\beta_M=10^{-15}$, the induced waves peak at $h^2\Omega_{{\rm GW},0}=1.4\times10^{-9}$ for the Gaussian bath and at $6.9\times10^{-10}$ and $2.9\times10^{-10}$ for $\K=0.1$ and $0.3$ (Fig.~\ref{fig:benchmark}). All three lie more than three orders of magnitude above the LISA power-law-integrated sensitivity at that frequency. At such a benchmark the background would be easy to detect, and the real task would be to interpret its amplitude, which differs between the three baths by factors of two and five.


\subsection{The induced-wave amplitude at a given collapse statistic}

The same result can be turned around. If we hold the collapse statistic fixed, a non-Gaussian bath reaches it with a smaller covariance, and therefore with weaker induced waves. Figure~\ref{fig:plane}a shows the ratio $\OGW/\OGW^{\rm G}=(\sigma/\sigma_{\rm G})^4$ needed to reach a given $\beta_M$. For $\beta_M=10^{-8}$, $10^{-12}$ and $10^{-16}$ it is $0.69$, $0.55$ and $0.46$ at $\K=0.1$, and $0.39$, $0.26$ and $0.19$ at $\K=0.3$. The rarer the black holes, the more the non-Gaussian bath gains, and at $\K=0.3$ the induced signal can be five times weaker than a Gaussian bath would require.

\subsection{Covariance amplitude versus statistics}
\label{sec:separating}

Figure~5b compares how changes in the covariance amplitude and the
excess kurtosis affect $\Omega_{\rm GW}$ and $\beta_M$. At fixed spectral shape, a change in the covariance rescales $\sigma^2$ by some factor $s$ and the induced waves by $s^2$, and so slides the point along a curve of fixed statistics. The higher cumulants instead move the point from one curve to another at fixed induced-wave amplitude. The collapse statistic alone cannot tell these two moves apart, but the pair of observables can. Start, for instance, from the covariance at which the Gaussian value is $10^{-12}$. Lowering the power by ten per cent lowers $\beta_M$ by a factor $10^{1.1}$ for a Gaussian bath, and a curvature kurtosis $\K=0.033$ brings it back. A twenty per cent reduction lowers $\beta_M$ by $10^{2.6}$ and needs $\K=0.072$ to compensate. The collapse statistic ends where it started, but the induced waves are weaker by factors of $0.81$ and $0.64$, so the two measurements together separate a change in the covariance amplitude from a change in the statistics.

\begin{figure*}[t]
\centering
\includegraphics[width=\textwidth]{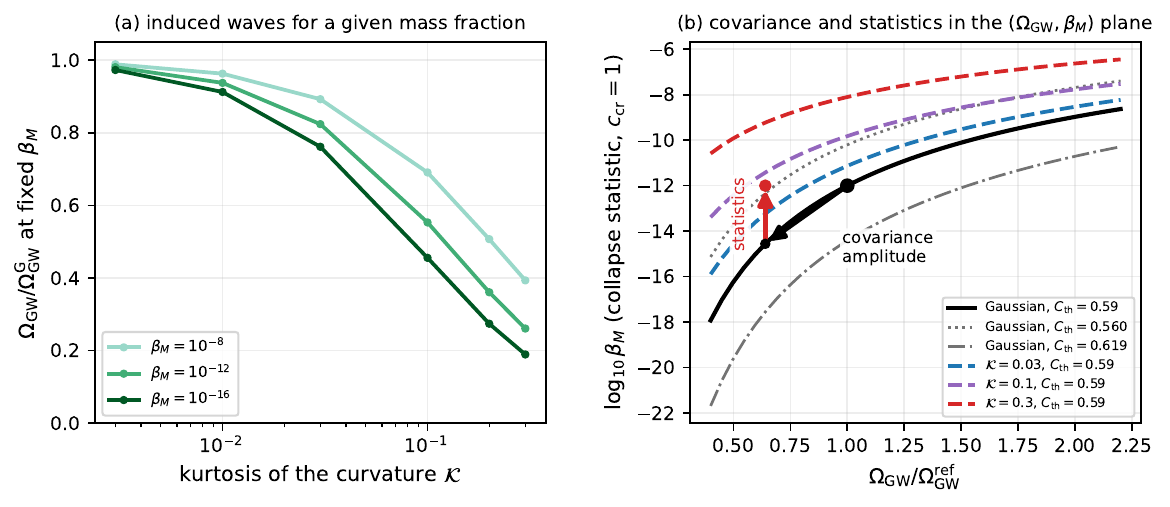}
\caption{(a) The induced-wave amplitude a non-Gaussian bath needs to reach a given collapse statistic, relative to what a Gaussian bath needs, as a function of the curvature kurtosis. The rarer the black holes, the larger the saving. (b) Covariance amplitude and statistics in the plane of the two observables, for the spectrum of Eq.~(\ref{eq:lognormal}). The vertical axis is the collapse statistic of Eq.~(\ref{eq:beta}) with $c_{\rm cr}=1$, and the horizontal axis is the induced-wave amplitude, normalised to the Gaussian value that gives $\beta_M=10^{-12}$. The solid black curve is the Gaussian bath with $\Cth=0.59$, and the grey curves are Gaussian baths with $\Cth$ shifted by $\pm5\%$. The coloured dashed curves are compound-Poisson baths with $\Cth=0.59$ and $\K=0.03$, $0.1$ and $0.3$. The arrows show the two moves discussed in the text: a $20\%$ drop in the covariance amplitude, one way in which memory can act, slides the point down the Gaussian curve, and a change in the statistics then lifts it back to the original $\beta_M$ at the new, lower amplitude.}
\label{fig:plane}
\end{figure*}

Memory is one natural way to change the covariance amplitude. In the freeze-out prescription of Ref.~\cite{gangopadhyay2026a} the power is suppressed by $1/(1+\eta_c^2)$, with $\eta_c$ the correlation time in units of the inverse response frequency, so the ten and twenty per cent reductions above correspond to $\eta_c\simeq0.33$ and $0.5$. Deriving a correlation-time-dependent response and noise that satisfy the fluctuation--dissipation relation within the present model is the natural next step. Other changes push the point the same way, among them changes in the potential, the dissipation, the temperature, the response or the normalisation of the source, as well as a quadratic nonlinearity of the size considered in Sec.~\ref{sec:robust}.  What the plane separates, then, is a change in the covariance amplitude from a change in the statistics, with memory as one route to the former. A change in memory can also reshape the spectrum, which moves the curves themselves, and a longer correlation time reduces the number of independent kicks sampled during freeze-out and can raise $\K$; the plane then constrains the combination, with $\K$ acting as an effective parameter.
 
Once the model is specified, a joint measurement of the induced-wave amplitude and the black-hole abundance constrains the bath. For a given spectral shape, mark law, response kernel, threshold and formation prescription, the two measurements fix a point in Fig.~\ref{fig:plane}b, and the curve through that point fixes $\K$. A point above the Gaussian curve would rule out that particular Gaussian model, while placing the non-Gaussianity in the bath itself would take more, since any change in the inputs also moves the curves. The quantity that matters most for such an inference is the collapse threshold. At the reference covariance where the Gaussian value is $10^{-12}$, lowering $\Cth$ by five and ten per cent raises $\log_{10}\beta_M$ by $1.8$ and $3.2$, while $\K=0.03$ and $0.1$ raise it by $0.87$ and $2.18$. At that point $\K=0.03$ has the same effect as a $2.3$ per cent lower threshold, and $\K=0.1$ as a $6.2$ per cent lower one. With the other model inputs fixed, an independent determination of
$C_{\rm th}$ at this precision would allow the two observables to
constrain $\mathcal{K}$. An additional observable could help if its
dependence on $\mathcal{K}$ and $C_{\rm th}$ differs from that of
$\beta_M$.


\section{Relation to other non-Gaussian constructions}
\label{sec:other}

It helps to place the model next to the two non-Gaussian constructions most often used in the PBH literature. The first is local non-Gaussianity, $\zeta=\zeta_g+\frac{3}{5}f_{\rm NL}\zeta_g^2+\dots$ \cite{maldacena2003,bartolo2004}, together with the trispectrum parameters $\tau_{\rm NL}$ and $g_{\rm NL}$ \cite{byrnes2006,suyama2008} that Planck constrains on CMB scales \cite{planck2020ng}. At leading order it acts on the abundance through the skewness, and the sign of $f_{\rm NL}$ decides whether black holes are enhanced or suppressed \cite{byrnes2012,young2013,franciolini2018}. The second is quantum diffusion in stochastic inflation, which produces exponential tails that can dominate the abundance \cite{vennin2015,ezquiaga2020,pattison2021}. The marked-Poisson bath differs from both in instructive ways. Its odd cumulants vanish by construction, so its leading effect is the kurtosis rather than the skewness. Its even cumulants are not free parameters but follow the hierarchy of Eq.~(\ref{eq:hierarchy}), set by the mark distribution and the response kernel. Its far tail, controlled by the bounded marks through the cumulant generating function, falls off faster than an exponential but more slowly than a Gaussian. And because its trispectrum depends on each external momentum separately, the leading induced waves do not see it at all.

The symmetric marks remove the skewness of the source, but the nonlinear dynamics of warm inflation can add a bispectrum of its own \cite{moss2007,bastero2014}. The explicit computation of Mirbabayi and Gruzinov \cite{Mirbabayi2022}
shows that this bispectrum is peaked in the squeezed and folded
configurations and carries information about the microphysics of the
heating. Their result reinforces the division of labour advcated in Sec.~\ref{sec:spectrum} for a symmetric source:
the bispectrum controls the skewness and is the natural target for CMB
templates, while in the model studied here the kurtosis controls the far tail and is the natural target for PBHs. The two statistics are not competitors but probes of
different regions of the same distribution, and the induced
gravitational waves, being blind to the kurtosis at leading order,
single out the PBH channel as the one that can read it. The bispectrum itself is captured at leading order by a quadratic term,
$\zeta=\zeta_L+\tfrac{3}{5}f_{\mathrm{NL}}(\zeta_L^2-\sigma^2)$.
Since our linear field has $\langle\zeta_L^4\rangle=(3+\kappa)\sigma^4$,
the resulting skewness is:
\begin{equation}
\mathcal S_3\equiv\frac{\langle\zeta^3\rangle}{\sigma^3}\simeq3.6\,f_{\rm NL}\,\sigma\left(1+\frac{\K}{2}\right),
\label{eq:skewness}
\end{equation}
which reduces to the familiar $3.6\,f_{\rm NL}\sigma$ when $\zeta_L$ is Gaussian. At the covariances used here this is about $0.3\,f_{\rm NL}$, and it acts on the tail with the opposite parity to $\K$. The example of Sec.~\ref{sec:robust} shows what such a term does in practice: it mainly shifts the threshold, leaves the $\K$-enhancement at a fixed reference value of $\beta_M$ almost unchanged, and alters the covariance needed to reach that value. At second order, $f_{\rm NL}^2$ terms also feed the induced waves directly \cite{cai2019,adshead2021}, so the blindness of the induced waves to the statistics is a property of the symmetric, linear family studied here. The calculation treats the event density, mark distribution and response
kernel as inputs. Applying it to a specific warm-inflation model requires
a microscopic calculation of these quantities.
\section{Conclusions}
\label{sec:conclusions}

In warm inflation the primordial fluctuations have a physical source: the random force exerted by a radiation bath. This force has two distinct properties. Its covariance carries the memory of the bath, and its statistics describe how the strengths of the individual kicks are distributed. In this paper we followed both through to primordial black holes and to the induced gravitational-wave background, using a marked-Poisson source with a linear response as a concrete and fully calculable example. Because every cumulant of this source is known in closed form, the two properties can be varied independently and their effects traced exactly.

The two observables respond in very different ways. For independent, isotropic events the covariance fixes the curvature power spectrum and with it the leading induced-wave signal, while the angular selection rule of Ref.~\cite{garciasaenz2023} removes the connected trispectrum at that order. The higher cumulants are therefore invisible to the induced waves, yet they control the tail from which black holes form. For a narrow log-normal spectrum, a curvature kurtosis between $0.1$ and $0.3$ raises the mass-weighted collapse statistic at fixed induced-wave amplitude by a factor between $10^{0.9}$ and $10^{6.2}$ over Gaussian values from $10^{-8}$ to $10^{-16}$. Put the other way, the same level of black-hole formation can be reached with induced waves up to $5.3$ times weaker. The effect is robust. Counting maxima instead of points lowers the exponents by $0.2$ to $0.6$, a spectrum three times broader by $0.4$ to $0.6$, and profile scatter, a shape-dependent threshold, the time profile of the response, the sign of the kernel and a mild quadratic nonlinearity matter less still; none of them removes the enhancement. The model also shows clearly what sets the size of the effect. Beyond $\K$ itself, the enhancement depends on the response kernel through the window factors $R_n$ and on the shape of the marks, and $\K$ records the event rate as well as the marks, with the values considered here corresponding to thousands of events per bath cell.

These results give the two observables complementary roles. Within a specified model, a change in the covariance amplitude, of which memory is one source, moves both, while the kurtosis moves only the black holes. An induced-wave bound therefore constrains the black-hole abundance only once the statistics of the source are known, and a joint measurement can pin down $\K$. A benchmark with asteroid-mass black holes shows that this is a practical question: the induced signal would lie well above the LISA sensitivity, so the task would be to read its amplitude correctly rather than to detect it. The inference rests on the mark law, the kernel and the formation prescription, and most sharply on the collapse threshold, since at the reference covariance $\K=0.1$ has the same effect as a $6.2$ per cent change in $\Cth$; it also responds to any change in the covariance amplitude, not to memory alone. These points mark the natural next steps: a microscopic calculation of the higher cumulants and of the response of the bath, a peak-based abundance with the full mass function, and a threshold for the relevant profiles at the precision quoted above. All three are well-defined calculations. The tools developed here carry over directly to a bath whose cumulants and response are derived rather than assumed, and with those inputs this example would become a quantitative statement about warm inflation itself.

\begin{acknowledgments}
M.R.G.\ thanks the OMEG Institute, Soongsil University, for its hospitality during the visit in which this work was begun.
\end{acknowledgments}

\end{document}